# Security in Cryptocurrency

Chelsea Medina [1], Lily Shaw [2], Dissy Vargas [3], Sundar Krishnan [4]

Angelo State University, San Angelo, United States

[1] cmedina15@angelo.edu, [2] mshaw7@angelo.edu , [3] dvargas10@angelo.edu , [4] skrishnan@angelo.edu

***Abstract*— This paper discusses the mechanisms of cryptocurrency, the idea of using security in the system, and the popularity of it. To begin, the authors provide a background on cryptocurrency and how it works. The authors understand that while most people may be familiar with the concept, they may not know how it works. Next, the authors discuss the security of cryptocurrency in-depth within the paper. The authors also provide examples of attacks on cryptocurrency systems to show the vulnerabilities within the system. Lastly, the authors discuss the popularity of the system to further express the need for security in cryptocurrency.**

## I. Mechanishms of CryptoCurrency

### A. *Maintaining the Integrity of the Specifications*

Cryptocurrency security currently relies primarily on blockchain technology, and the consensus features offered by a decentralized currency and authority as a security measure for preventing DOS, double spending attacks, and exploits. The distributed nature of the ledger to track spending and mining on the network ensures that the data being received can be trusted and verified by multiple sources. These immutable security features of a blockchain system have been a great boon to the proliferation of cryptocurrencies. Cryptocurrencies rely on asymmetric cryptography to ensure the validity and security of entries on the blockchain and the equivalent transactions. Security on the networks of these currencies must be maintained to compete in the marketplace.

Asymmetric encryption is the backbone of security, allowing for a public key and private key pair to be used for signing or verification of transactions and wallet creation. The private key can be used by the user to access their account and make transactions using their tokens. The public key can be used to determine the validity of these transactions before their entry on the blockchain. Once a transaction is made, it is sent to a memory pool to be included in the next block in the blockchain. All the miners generating the blocks review transactions for ownership and transaction validity using the publicly available key. Once verified and attached to the blockchain, the ledger reflects the transaction, and the updated ledger is distributed to the nodes worldwide. Primarily the ledger used is the one with the most entries, to ensure the chain is the most up-to-date [10].

Blockchain as a concept relies upon blocks of data being stored together and having an immutable characteristic attached to them. The data blocks contain strings and the signatures that they hash too. The immutable quality of these blocks leads them to be highly resistant to tampering, malicious or not. Blockchains in cryptocurrencies are available publicly, as is the software to check the validity of the blockchain at any given point. The hashing algorithm for Bitcoin, as an example, is SHA-256, which generates an almost unique signature for the text contained inside the block. These blockchain ledgers allow for security through mass checking of the ledger to determine the legitimacy of a copy.

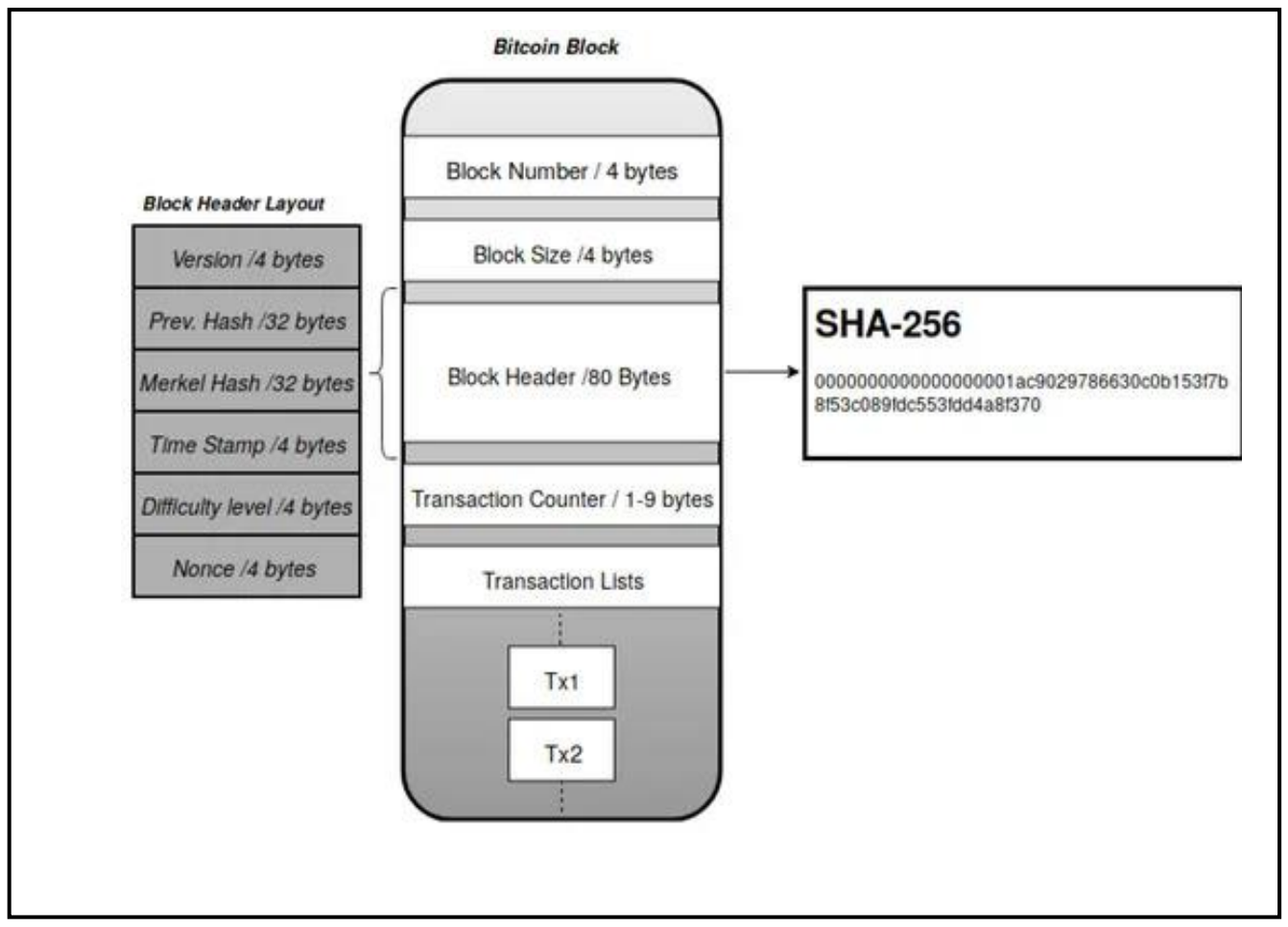

Figure 1: SHA-256 Hashing Algortihm [9]**.** Example of a block on the Bitcoin Blockchain

### B. *Proof of Work vs. Proof of Stake*

Consensus algorithms vary from currency to currency, many use either Proof of Work or Proof of Stake. Proof of work as a concept allows for competition in mining new blocks to add to the blockchain, thereby processing new transactions. Proof of Stake on the other hand relies on random selection of a miner and subsequently rewards that miner with a transaction fee for adding a new block. Both are effective at scale yet have their drawbacks. Proof of Work is vulnerable to consensus attacks where most of the hashing power is captured in some way. Proof of Stake however is vulnerable to whomever controls a large majority of the coins or tokens. Both have their potential issues, yet there are mechanisms to protect from them.

Issues with the Proof of Work model arise from mass capture of hashing power and the algorithms behind them, there are a few types of attacks such as the 51% attack and the 34% attack. Both rely on steady capture of hashing power to gain control of cryptocurrency mining operations and ledger operations as well. The primary concern is the concerted effort of many malicious actors could render a currency insolvent and unusable in a matter of days perhaps. The protections against this are primarily focused on devaluing blockchains that were delayed in their submission and notary nodes for verification. Both strategies have issues, but combined they can be a force to prevent capture of the blockchain and block double-spending attacks [9].

The Proof of Stake model is vulnerable to bribe attacks and a different variety of 51% attacks. Bribe attacks occur with an alternative chain being built, and then being pushed out to the network at large. Due to the randomized nature of the block creation, there is a much higher chance of the illicit blockchain being longer, and therefore considered more legitimate, than the correct blockchain. The Proof of Stake 51% attack relies on capturing the majority of the coins as opposed to hashing power. Because Proof of Stake depends on ownership of coins to process transactions, capturing a majority can allow for the creation of illicit blocks in the blockchain. The prevention of this uses similar concepts to the Proof of Work model. Devaluing blockchains that were not submitted immediately. A different method though occurs when the stakes of all nodes are equalized for deciding which ledger to choose, to prevent one single stakeholder from seizing control of an entire network [8].

### C. *User Security*

User security in some mainline cryptocurrencies and emerging ones as well is the idea of a main wallet and a sub-wallet to reduce account capture risks. The threat of account theft is mitigated by having a subwallet used for everyday spending and the main account acts as a savings account. The subwallet should experience more frequent use, putting it at much greater risk than the main wallet. These security controls must be implemented industry-wide to mitigate the threat to users of these currencies, due to the lack of FDIC insurance and regulations [7].

Other security vulnerabilities of cryptocurrencies stem from the lack of stabilization in the markets. Due to the nature of how crypto currencies are bought and sold, pump-and-dump schemes abound on crypto networks of all sizes. These schemes cause inflation of the value of a coin or token by incentivizing others to buy into the network. Once the value of a coin is high enough, the group organizing the scheme sells off their tokens, causing the valuation of the currency to drop sharply. Many cryptocurrencies and exchanges lack any protection from this kind of attack. These attacks can lead to a devastating loss of trust in the cryptocurrencies affected, setting back the progress being made by a network group. There are currently minimal ways to stop an attack or scheme like this from happening on a network. The only protection is in the users themselves, who must be wary of schemes like this.

Due to the wide variety of network sizes and lack of federal regulations surrounding cryptocurrencies, the security mechanisms lack any unified implementation or efficacy. Smaller networks remain very vulnerable to blockchain capture due to a relatively small number of hashing power needed to be captured. Larger networks find themselves less vulnerable to the hashing types of attacks, yet exchanges remain the bottleneck in the security of these currencies. The lack of a unified front on both exchanges and blockchains leads the door open to threats at every level and angle.

### D. *Security Solutions*

A possible solution to cryptocurrencies' security issues could be government intervention in the form of security standards required for all facets of the industry. While regulation goes against the ideals of cryptocurrency and most users, to attain widespread, consistent usage security must be at the front of mind for cryptocurrency developers. Government regulation is uncomfortable for most emerging industries yet, in the case of currencies it is a requirement for user safety and currency stability. Regulations abound for all other types of assets, both for security and exchanges, crypto should be treated no differently.

## II. ATTACKS IN CRYPTOCURRENCY

### A. *Common Attacks*

Cryptocurrency replaces the way we bank by not being regulated by government or bank authorities. Without anyone to make sure nothing suspicious happens with their digital assets, anything can happen. The most common attacks reported by Coin Base, which is a platform for cryptocurrency users, are account takeovers (ATO) and SIM-swaps. ATO attacks are when someone gains access to your login credentials and can login to your account to commit fraudulent activities [12]. A SIM-swap attack is when bad actors contact your phone carrier pretending to be you and persuade the employee on call to redirect your cell service to another one by changing the SIM card number. When the SIM card number is changed the bad actor will then receive all calls and text messages that are sent to you. These text messages can sometimes have two-factor authentication codes that a threat actor could use to authenticate themselves and get into any account a person has. Both have been very common attacks for anyone who has cryptocurrency assets. Various news outlets have reported on the number of crypto assets stolen in recent times.

### B. *Modern Day*

For 2022, it was estimated that more than three billion in cryptocurrency has been stolen by threat actors [1]. Even with large cryptocurrency exchange companies that say they are secured they are still being hacked and robbed of assets. Having security for crypto is still beneficial and can reduce the number of attacks. Banks still get robbed but having security guards acts as a deterrent for bad guys.

## III. Cryptocurrency popularity

Cryptocurrency has slowly been gaining traction since its initial formation back in the 1900s. It originated in the 1990s, but has recently been rising in popularity. These past few years have shown remarkable growth in the cryptocurrency field. With that comes its challenges and questions surrounding cryptocurrency.

### A. Adjusting to Cryptocurrency

The way that cryptocurrency works allows users to be in control of several aspects of how it works. Instead of having companies like Visa or Paypal, there are networks that users can connect on. When done this way, the transaction is now between two people with no third party. Some of the more popular networks include Bitcoin, Litecoin, and Ethereum [6].

Bitcoin, in particular, has become a household name in the past few years. While many people may not know about the specifics of cryptocurrency, at least 86% of Americans have heard of the relatively new system. So, we can safely assume that at this point, most Americans have heard of cryptocurrency in some form. Initially, it was not this popular, but with the recent popularity rise it's become a major player in the finance world.

As a result, what has happened and is still ongoing is the economic adjustment to make space for cryptocurrency. To put it in more specific terms, Bitcoin (a successful cryptocurrency system) alone accounts for $6 billion in online transactions daily. This is a significant amount of transactions and shows a steep increase from the previous transactions.

### B. Using Cryptocurrency

For the year 2021, daily transactions reached up to nearly 400,000. User wise there were 68 million wallet users for the year 2021 in February [2]. Wallets are a term used in cryptocurrency to refer to where a user stores their personal data. They are used the same way as bank accounts or an actual physical wallet are used, to store money. Since cryptocurrency deals with virtual money they store personal data in the case of the digital wallet. Through that wallet, users can send and receive currency.

It seems like an easy enough process. New users are joining every day and there has definitively been an increase from when cryptocurrency was first introduced. Of course, with a system like this comes its own advantages and disadvantages.

### C. Advantages & Disadvantages

An advantage to using cryptocurrency is the ease with which it can be used. The process of creating a wallet and using it can be easier than opening a bank account in some places. Another advantage is the transparency and anonymity of the system. While all transaction history is saved, the system manages to be completely anonymous at the same time. Users can create Bitcoin addresses without any reference to what can be considered personal information such as name or address. Additionally, when making transactions, it does not require users to enter personal data. Normally when using credit cards or other services users are required to fill out personal information to use them, but this is different with cryptocurrency. There are public and private keys that are used as explained earlier [3].

While there are some definite advantages to using cryptocurrency, it also comes with some steep disadvantages. The system's biggest drawback is its constant risk of hacking. Cryptocurrency is not backed by any banks and there is typically no insurance leading it to become a hotspot for hackers and scams. It means users have to be careful when they are making transactions, which can make it lose appeal with certain people.

### D. Attack Vulnerability

A negative of this system is its vulnerability to scams leaves many users unsure and unwilling to participate due to the risk of losing money. For example, just this month two individuals were arrested for a $575 million cryptocurrency fraud. They used their cryptocurrency mining service called HashFlare to trick victims into entering fraudulent contracts. This incident happened from 2015 to 2019, so for about four years, they were convincing users to enter these contracts and not paying off those who had already invested [11]. This is one of the many examples of the problems with cryptocurrency. When working with any sort of money exchange, there is a high risk involved, but it becomes even more complicated when you're using cryptocurrency.

### E. Price Inconsistency

Another issue is the inconsistency of prices. This can be seen when examining the history of Bitcoin. The original investment cost for Bitcoin was next to nothing. The value of a Bitcoin was $0 back in 2009. It has now risen to opening at around $48,000 for 2022. This inconsistent pricing has caused uncertainty among investors. Although, it should be noted that it's expected to continue growing up to another $10,000 in the next few years [4].

Table-1 further shows the value discrepancy not only over time, but also between platforms. As shown, there have been substantial changes that have occurred. In the table there are three row values. The "Start Value" in the table is the value that the platform starting trading at. There are three platforms show, Bitcoin, Ethereum, and Litecoin. There are others, but these tend to be the most popular ones. Notice that there has been a notable drop from the opening value to today's value. This could be attributed to several factors.

### F. Drop In Popularity

All the previous disadvantages could have led to some people losing interest in using cryptocurrency. They may be part of attacks that occur and be discouraged from participating or even seeing these attacks occur could be a reason for not joining. This is one of cryptocurrency's largest problems. Additionally, the risk of investing with the volatile nature of cryptocurrency can be a deal breaker for others.

This drop in popularity has not largely hindered the new system as there is still a push for the new transactional system. Many countries are now beginning to accept crypto as a valid form of payment/exchange, meaning it could have a permanent place as a transactional system.

<table>
<tr><th rowspan="2">Values</th><th colspan="3">Cryptocurrency Platforms</th></tr>
<tr><th>Bitcoin</th><th>Ethereum</th><th>Litecoin</th></tr>
<tr><td>Value in Oct 2023</td><td>$26,870.44</td><td>$1,547.29</td><td>$61.54</td></tr>
<tr><td>Start Value</td><td>$0.0008<br>(In 2010)</td><td>$0.31<br>(In 2015)</td><td>$4.30<br>(In 2013)</td></tr>
</table>

Table 1: Values in Cryptocurrency Platorms [4].

## IV. Conclusion

With the continued growth of cryptocurrency, society will continue to make adjustments. While its popularity may fluctuate from time to time, it seems that for now, the system is here to stay. Thus, there is importance in having the appropriate security for it. Certain vulnerabilities make it a good target for attacks from hackers. The lack of stabilization within the industry and the lack of regulations leaves the system open for attacks, which have been on the rise as cryptocurrency gains popularity. One recommendation the authors recommend is government oversight to safeguard financial investments of it's citizens from financial irregularities. It does appear to be against what cryptocurrency originally was created for, but the lack of security standards can be damaging to users who get involved in attacks.